\documentclass[%
 reprint,
superscriptaddress,
longbibliography,
amsmath,amssymb,
]{revtex4-1}

\usepackage{bbold}
\usepackage{mathptmx}
\usepackage{subfig}
\usepackage{psfrag,graphicx}
\usepackage{dcolumn}
\usepackage{amsmath,amssymb}
\usepackage{bm}
\usepackage{color}
\usepackage{latexsym}
\usepackage{epstopdf}
\usepackage{color}
\usepackage[english]{babel}
\usepackage{latexsym}
\usepackage{psfrag,graphicx}
\usepackage{amsmath}
\usepackage{amssymb}
\usepackage{amsfonts}
\usepackage{bm}
\usepackage{natbib}
\usepackage{epstopdf}
\usepackage{appendix}
\usepackage{rotating}
\usepackage[english]{babel}
\usepackage{aeguill}
\usepackage{ulem}
\usepackage[justification=justified]{caption}

\definecolor{mygrey}{gray}{0.35}
\definecolor{myblue}{rgb}{0.2,0.2,0.8}
\definecolor{myzard}{cmyk}{0,0,0.05,0}
\definecolor{mywhite}{rgb}{1,1,1}
\definecolor{mywhite}{rgb}{1,1,1}
\definecolor{myred}{rgb}{1,0.,0.3}

\usepackage[colorlinks=true,citecolor=myblue,linkcolor=myblue]{hyperref}

\def\ba{\begin{align}}
\def\enda{\end{align}}
\def\bi{\begin{itemize}}
\def\ei{\end{itemize}}

\def\be{\begin{equation}}
\def\ee{\end{equation}}
\def\bea{\begin{eqnarray}}
\def\eea{\end{eqnarray}}
\def\bse{\begin{subequations}}
\def\ese{\end{subequations}}

\newcommand{\ket}[1]{|{#1}\rangle}                       % ket
\newcommand{\average}[1]{\langle {#1} \rangle}           % media < >

\newcommand{\ketbra}[2]{\left\vert#1\right\rangle\left\langle#2\right\vert}

\newcommand{\Ignore}[1]{ }

\def\i{\text{i}}

\begin{document}

\preprint{APS/123-QED}

\title{
{\color{black}
Jaynes-Cummings dynamics in strong coupling for many-interacting-qubit quantum Rabi models
}
\Ignore{\\
Definition of strong/weak coupling regimes for many-interacting-spin quantum Rabi models}
}

\author{Roberto Grimaudo}
\address{Department of Physics and Astronomy ``E. Majorana", University of Catania, Via S. Sofia, 64 I-95123 Catania, Italy}
\address{INFN Sez. Catania, 95123 Catania, Italy}

\author{Sagnik Chakraborty}
\address{Dipartimento di Ingegneria, Universit\`{a} degli Studi di Palermo, Viale delle Scienze, 90128 Palermo, Italy}

\author{Rosario Lo Franco}
\address{Dipartimento di Ingegneria, Universit\`{a} degli Studi di Palermo, Viale delle Scienze, 90128 Palermo, Italy}

\author{Giuseppe Falci}
\address{Department of Physics and Astronomy ``E. Majorana", University of Catania, Via S. Sofia, 64 I-95123 Catania, Italy}
\address{CNR-IMM, UoS Universit\`a, 95123, Catania, Italy}
\address{INFN Sez. Catania, 95123 Catania, Italy}

\date{\today}

\begin{abstract}

{\color{black}
The present work focuses on the strong/weak interaction of many-body spin-systems with a cavity mode.
It introduces the necessity of redefining the physical conditions determining the strong/weak coupling regime in those systems.
In more complex systems, the effective coupling emerging from the collective dynamics may differ indeed from the actual coupling of each individual subsystem with the bosonic field.
This is shown by highlighting some counter-intuitive dynamical effects properly related to the coupling regime: a Jaynes-Cummings dynamics emerging although a strong interaction is present.
The universality of this result is demonstrated through the analysis of three distinct systems: a two-qubit, a two-qutrit, and an $N$-qubit chain quantum Rabi models.
}

\end{abstract}

\pacs{ 75.78.-n; 75.30.Et; 75.10.Jm; 71.70.Gm; 05.40.Ca; 03.65.Aa; 03.65.Sq}

\keywords{Suggested keywords}

\maketitle

The strength of the coupling between light and matter plays a critical role in determining the dynamics of systems involving these two fundamental components.
The simplest theoretical framework describing the interaction between a two-level atom and a quantized field mode is the quantum Rabi model (QRM).
This model can be implemented on various platforms, such as cavity quantum electrodynamics (cavity QED) and circuit QED \cite{Forn19}, and semiconductors-based technologies such as quantum dots.

In the weak coupling regime, the light-matter interaction strength is much smaller than the natural frequencies of the two-level system and the field mode.
Under these conditions, counterrotating terms in the QRM can be neglected, simplifying the QRM to the Jaynes-Cummings model (JCM) \cite{Larson_book}.
The coupling regime then determines the form of the Hamiltonian, which in turn affects the symmetries and conserved quantities of the system.
In fact, while the QRM exhibits a $Z(2)$ symmetry, related to the conservation of the parity of the number of excitations, the JCM is characterized by a $U(1)$ symmetry, related to the conservation of the total number of excitations, which divides the Hilbert space into an infinite set of two-dimensional invariant subspaces.
When the coupling strength becomes a significant fraction of the natural frequencies of the noninteracting parts, the ultra-strong coupling regime emerges.
Here, counter-rotating terms can no longer be neglected, leading to notable dynamical effects such as the Bloch-Siegert shift \cite{Ahmad_74, Hannaford_73, Yan15}.
In the deep-strong coupling regime, where the coupling strength exceeds the natural frequencies of the system, the interaction terms become non-perturbative, resulting in further distinct dynamical behaviors \cite{Yoshihara17}.

Such a variety of dynamical regimes is expected also for more complex systems, of course.
However, for many-body systems these conditions can differ significantly.
Identifying the physical conditions that define these coupling regimes for such systems is essential for optimizing light-matter systems in applications for quantum technologies like quantum information processing, quantum computation, and quantum sensing. 

This study shows that in many-spin systems interacting with a quantized field, the coupling regime criteria differ from those of the single-qubit QRM. This arises from invariant subspaces, where dynamics is governed by effective Hamiltonians with redefined parameters.

Analyzing cases such as the two-qubit QRM, the two-qutrit QRM, and an $N$-qubit chain QRM we first find that the coupling regime depends on the ratio of spin-mode to spin-spin coupling, rather than the conventional spin-mode to spin-frequency ratio.
Further, collective dynamics can lead to unexpected coupling behaviors, where collapse and revival can occur in strong coupling, and effective strong coupling can emerge from weak individual interactions.

These results highlight that in many body systems, the definition of the coupling regime requires careful consideration, since the symmetries exhibited by the Hamiltonian as well as the initial state of the system significantly influence the effective strength of the interaction.

\textit{Two-qubit QRM.}
Let us focus on the following Hamiltonian model (in units of $\hbar$):
\begin{equation} \label{Hamiltonian}
\begin{aligned}
{H} = &
\omega a^\dagger a
+ \epsilon_1 \sigma_1^z + \epsilon_2 \sigma_2^z
+ \gamma\sigma_{1}^{x}\sigma_{2}^{x}
+ (\lambda_1 \sigma_1^z + \lambda_2 \sigma_2^z) (a +a^\dagger).
\end{aligned}
\end{equation}
It describes two distinguishable interacting qubits, characterized by distinct splitting energies ($\epsilon_1$ and $\epsilon_2$), and coupled to a single bosonic field mode with characteristic frequency $\omega$.
The coupling constants $\lambda_i$ ($i=1,2$) describe the interaction between the $i$-th qubit and the bosonic mode, while $\gamma$ represents the direct coupling between the two qubits.
The Pauli operators for the qubits are denoted as $\sigma_{k}^{l}$ ($k=1,2$, $l=x,y,z$), and the bosonic annihilation and creation operators are represented by $a$ and $a^\dagger$, respectively.

Basing on the existence of the constant of motion $\sigma_{1}^{z}\sigma_{2}^{z}$, we can write the Hamiltonian as $H = H_{a} \oplus H_{b}$, with \cite{GdCMSV}
\begin{subequations} \label{H+ H-}
\begin{align}
{H}_{a} = &
\omega ~ a^\dagger a +
\epsilon_+ \sigma_{a}^z +
\gamma \sigma_{a}^{x} +
\lambda_+ \left( a^\dagger + a \right)  \sigma_{a}^z,
\\
{H}_b = &
\omega ~ a^\dagger a +
\epsilon_- \sigma_{b}^z +
\gamma \sigma_{b}^{x} +
\lambda_- \left( a^\dagger + a \right)  \sigma_{b}^z, \label{Hb}
\end{align}
\end{subequations}
with $\epsilon_\pm = \epsilon_1 \pm \epsilon_2$, $\lambda_\pm = \lambda_1 \pm \lambda_2$, and $\sigma_{a/b}^l$ ($l=x,y,z$) are two-level Pauli operators.

The Hamiltonians $H_{a}$ and $H_{b}$ can be interpreted as effective descriptions of the dynamics of the two-qubit-mode system within the dynamically invariant subspaces $\mathcal{H}_{a}$ and $\mathcal{H}_{b}$, respectively.
The subspace $\mathcal{H}_{a}$ is spanned by ${ \ket{\uparrow\uparrow}, \ket{\downarrow\downarrow} } \otimes \{ \ket{n} \}_{n=0}^\infty$, while $\mathcal{H}_{b}$ is spanned by ${ \ket{\uparrow\downarrow}, \ket{\downarrow\uparrow} } \otimes \{ \ket{n} \}_{n=0}^\infty$. Here, $\sigma^z \ket{\uparrow} = +\ket{\uparrow}$, $\sigma^z \ket{\downarrow} = -\ket{\downarrow}$, and $a^\dagger a \ket{n} = n \ket{n}$ \cite{GdCMSV, GVSM}.

This framework implies that the two qubits collectively behave as an effective single two-level system within each invariant subspace.
The Hamiltonians $H_{a}$ and $H_{b}$ correspond to single-spin QRM Hamiltonians, where the qubit-qubit coupling $\gamma$ acts as the transverse field on the (fictitious) two-level systems.

It is important to note that while the field operators in $H_{a}$ and $H_{b}$ are written identically to those in $H$ for convenience, they are formally distinct.
Specifically, the operator $(a + a^\dagger)$ in $H_{a}$ and $H_{b}$ should be understood as $(a + a^\dagger) \otimes (\ketbra{\uparrow \uparrow}{\uparrow \uparrow} + \ketbra{\downarrow \downarrow}{\downarrow \downarrow})$ for $H_{a}$, and as $(a + a^\dagger) \otimes (\ketbra{\uparrow \downarrow}{\uparrow \downarrow} + \ketbra{\downarrow \uparrow}{\downarrow \uparrow})$ for $H_{b}$.

This decomposition simplifies the analysis of the original system of two interacting qubits, coupled to the same quantized field mode, by reducing it to the study of two independent effective problems related to two distinct single-spin QRMs.
Notably, this approach remains valid for Hamiltonian parameters that vary with time and across all qubit-mode coupling regimes (for a detailed classification of these coupling regimes, see Refs. \cite{Forn19, Kockum19, Xie17}).

\textit{JC subdynamics.}
Let us consider two equally biased ($\epsilon_1=\epsilon_2$) qubits, and strongly coupled to the field mode through an almost equal interaction strength, that is, $\lambda_1 \approx \lambda_2$.
In this case, on the one hand, the effective Hamiltonian $H_a$ turns out to be an anisotropic QRM \cite{Mao21}.
On the other hand, even if each qubit strongly interacts with the cavity mode, the effective coupling in $H_b$ results to be almost vanishing, $\lambda_- \approx 0$, and can be then considered very much smaller than $\omega$ and $\gamma$, namely $\lambda_- \ll \gamma,\omega$.
Therefore, the subdynamics of the two qubits restricted to the subspace $\mathcal{H}_b$, is characterized by a weak effective qubit-mode interaction, although the two qubits are in a strong coupling regime.

In this instance, we are then legitimated to treat the effective Hamiltonian $H_b$ by considering approximations typical of the weak coupling regime.
Therefore, by rotating $H_b$ of $\pi/2$ with respect to the $y$ direction, by unitarily transforming it through the operator $e^{i\omega(a^\dagger a + \sigma_b^z)/\hbar}$, and by neglecting the counter-rotating terms, the resulting effective Hamiltonian governing the (sub)dynamics under scrutiny acquires the JC form (see Supplemental Material):
\begin{equation} \label{H- new}
\begin{aligned}
\widetilde{H}_{b} = 
%\omega ~ a^\dagger a -
-\Delta \sigma_b^{z} + 
\lambda_-(a\sigma_b^+ + a^\dagger\sigma_b^-),
\end{aligned}
\end{equation}
where $\Delta = \gamma - \omega$ can be identified as an effective detuning.
However, such a detuning does not depend on the difference between the qubit and mode frequencies; rather, it is determined by the difference between the qubit-qubit coupling and the mode frequency.
This arises because, as previously emphasized, the qubit-qubit coupling acts as the effective splitting energy for the fictitious spin $b$.
Consequently, the resonance condition $\Delta=0$ for the two-qubit JCM under consideration is achieved when the characteristic frequency of the cavity is matched by the qubit-qubit coupling.
{
\color{black}
We stress that the derivation of the above JC Hamiltonian is fundamentally different from that presented in Ref. \cite{CattaneoJoP19}. In that work, the authors follow the standard procedure for deriving the master equation for two qubits coupled to a (common or separate) bath.
By contrast, our approach exploits the symmetries of the Hamiltonian, allowing for an easier and more direct derivation of the effective JC Hamiltonian governing the specific subdynamics of interest.
In particular, symmetry considerations reveal the existence of two invariant subspaces in which the two qubits effectively behave as a single two-level system.
Within one of these subspaces, and under suitable physical conditions on the Hamiltonian parameters, we show that the system can be described by an effective Hamiltonian corresponding to a single qubit coupled to the mode, exhibiting a JC form.
}

It is important to note that, for the physical system under consideration, also the definition of weak coupling differs from the conventional one.
Here, the critical factor is not only the ratio $\lambda/\omega$ but also the ratio $\lambda/\gamma$, which represents the relationship between the two couplings that characterize the system.
This contrasts with the usual focus on the ratio between the qubit-mode coupling and the qubit frequency, $\lambda/\epsilon$.

{\color{black}
We emphasize that, unlike most works in the literature \cite{ZhaoPra11, MazzolaPra09, SainzPra07, ShenPra19, BellomoPrl07, BellomoPra08, LoFrancoIJMP13, FicekPra06}, we do not assume a weak qubit–mode coupling from the outset, which would allow one to apply weak‑coupling approximations.
In those studies, the qubits are typically weakly coupled to a common bath, allowing to write a Jayne-Cummings Hamiltonian.
In Refs. \cite{HuangPra17, HuangPra20}, the authors instead render the counter-rotating terms approximately negligible by dynamically modulating the frequencies of both the qubits and the mode.
In our approach, by contrast, we work in the strong (up to ultra‑ or even deep‑strong) coupling regime.
Nonetheless, by analytically treating the Hamiltonian, we identify the physical conditions under which the qubit–mode interaction effectively results to be as if it the qubits were weakly coupled to the mode.
This then allows us to perform certain approximations and deriving the effective Jaynes-Cummings Hamiltonian.
Consequently, this leads to some counter‑intuitive effects discussed below: although both qubits are strongly coupled to the field mode, the resulting dynamics resemble those typically associated with a weak‑coupling regime.
}

\begin{figure}[htp]
\begin{center}
\subfloat[][]{\includegraphics[width=0.22\textwidth]{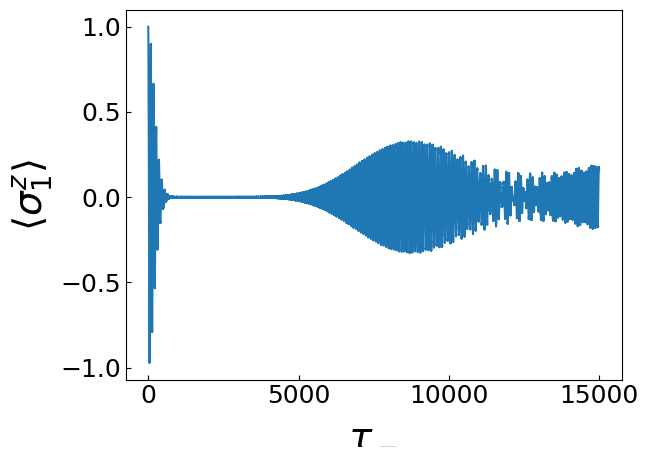}\label{fig: mv_z_an}}
\qquad
\subfloat[][]{\includegraphics[width=0.22\textwidth]{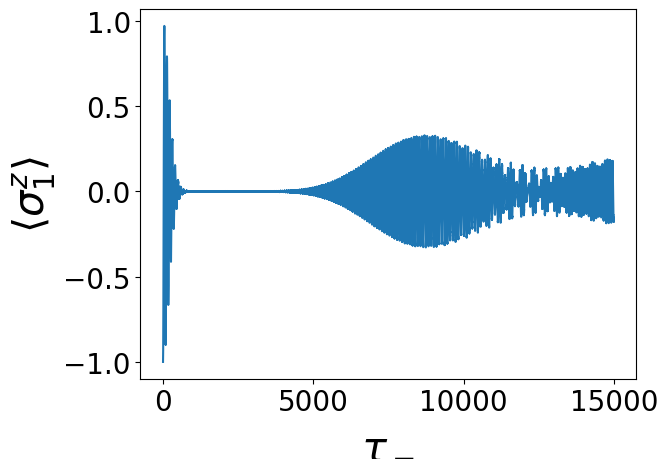}\label{fig: mv_z_num}}
\qquad
\subfloat[][]{\includegraphics[width=0.22\textwidth]{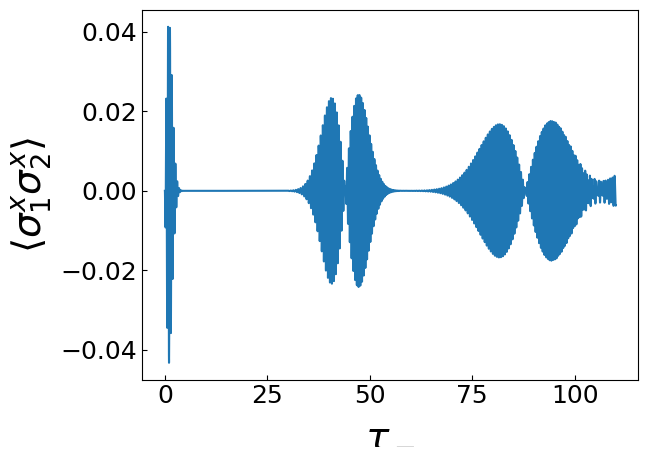}\label{fig: mv_xx_an}}
\qquad
\subfloat[][]{\includegraphics[width=0.22\textwidth]{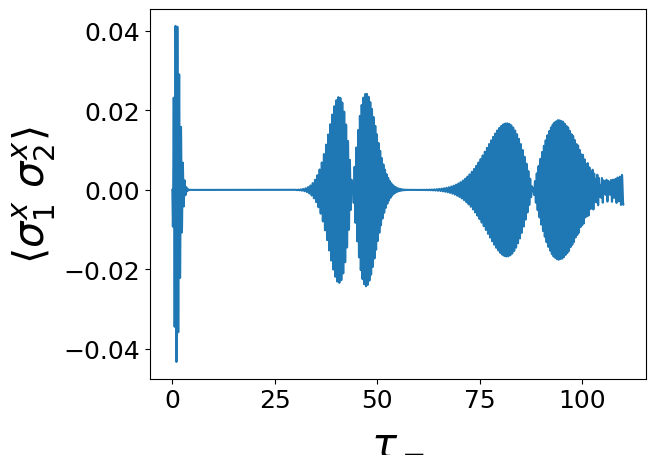}\label{fig: mv_xx_num}}
\captionsetup{justification=raggedright,format=plain,skip=4pt}%
\caption{ Time behavior, versus $\tau_-=\lambda_-t$, of: $\average{\sigma_1^z}$ [(a) analytical, (b) numerical], and $\average{\sigma_1^x\sigma_2^x}$ [(c) analytical, (d) numerical], when the system is initially prepared in $\ket{\psi (0)} = \ket{\uparrow \downarrow} \otimes \ket{\alpha}$, for $\Delta=0$ ($\gamma=\omega$: resonance), $\lambda_-/\omega=0.0001$ ($\lambda_1/\omega=0.8, ~ \lambda_2/\omega= 0.7999$), $\alpha=7$.}
\label{fig: mvs subdynamics b}
\end{center}
\end{figure}

Let us consider the resonance case, $\Delta=0$, when the system is initially prepared in the state
\begin{equation} \label{In State}
\ket{\psi (0)} = \ket{\uparrow \downarrow} \otimes \ket{\alpha},
\end{equation}
where $\ket{\alpha}=\sum_{n=0}^\infty c_n \ket{n}$, with $c_n = e^{-|\alpha|^2/2} {\alpha^n \over \sqrt{n!}}$, is a coherent (Glauber) state of the field, ($\ket{\alpha}=e^{\alpha a^\dagger - \alpha^*a}\ket{0} \equiv D(\alpha) \ket{0}$).
The mean value in time of the operators $\average{\sigma_j^z}=\average{\psi(t)|\sigma_j^z|\psi(t)}$ ($j=1,2$) ($\ket{\psi(t)}$ being the time evolved state of $\ket{\psi(0)}$ in Eq. \eqref{In State}) can be analytically derived (see Supplemental Material), turning out to be
\begin{equation} \label{eq: mv z an}
\begin{aligned}
 \average{\sigma_1^z} &= - \average{\sigma_2^z}= \\
 \sum_{n=0}^\infty [ & c_n^2 \cos(\sqrt{n+1} \tau_-) \cos(\sqrt{n} \tau_-) \\
 &+ c_{n+2} c_{n} \sin(\sqrt{n+2} \tau_-) \sin(\sqrt{n+1} \tau_-) ],
\end{aligned}
\end{equation}
with $\tau_- = \lambda_- t$.
Its time behavior according to Eq. \eqref{eq: mv z an} is shown in Fig. \ref{fig: mv_z_an} and perfectly agrees with the numerical simulation in Fig. \ref{fig: mv_z_num}.
A perfect accordance is also shown by
\begin{equation} \label{mv xx}
\begin{aligned}
    & \average{\psi(t)|\sigma_1^x \sigma_2^x|\psi(t)} = \\
    & \sum_{n=0}^\infty \{ c_n^2 [\cos^2(\sqrt{n+1}\tau_-) - \cos^2(\sqrt{n}\tau_-)] \\
    & \hspace{0.5 cm} + (c_{n+1}^2 - c_{n}^2) \sin^2(\sqrt{n+1}\tau_-) \},
\end{aligned}
\end{equation}
plotted in Figs. \ref{fig: mv_xx_an} (theoretical expectation) and \ref{fig: mv_xx_num} (numerical result).
In both cases the phenomenon of collapse and revivals, typical of a JCM dynamics, is clearly visible.
{\color{black}
This confirms the establishment of an effective weak coupling regime for the collective dynamics (ruled by $\widetilde{H}_{b}$ in Eq. \eqref{H- new}), although the single qubits are strongly coupled to the field mode.
Such an unexpected and counter-intuitive dynamical effect has never highlighted in the previous works since in those cases a weak qubit-mode interaction has been considered since the beginning.
}
\Ignore{
The quantities considered above exhibit collapses and revivals characterized by the corresponding characteristic times \cite{Larson_book}
\begin{equation}
    t_c = {1 \over 2\lambda_-},
    \quad
    t_r = {2 \pi m \sqrt{\bar{n}} \over \lambda_-},
\end{equation}
respectively, when $\bar{n} = \average{\hat{n}} = |\alpha|^2 \gg 1$ (in order to expand the Rabi frequency \cite{Larson_book}).
The $m$ in the above expression for $t_r$ indicates the $m$-th revival.
Both expressions, scale as $\lambda^{-1}$, implying that the closer the values of $\lambda_1$ and $\lambda_2$ are (and thus the smaller $\lambda_-$ is), the longer the collapse and revival times become.
}

\begin{figure}[htp]
\begin{center}
\subfloat[][]{\includegraphics[width=0.25\textwidth]{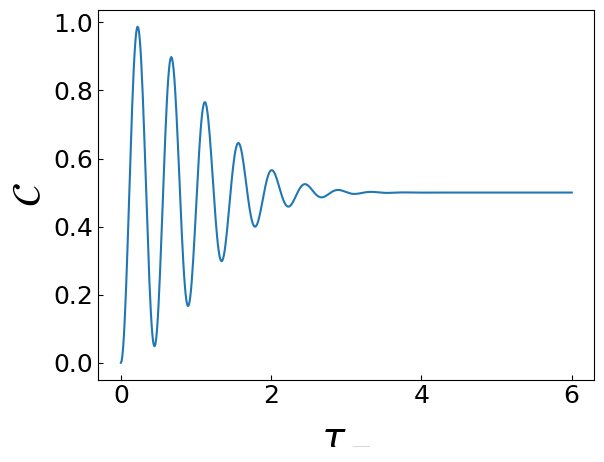}\label{fig: Conc_JCM}}
\caption{Time behavior of the concurrence $\mathcal{C}(\tau_-)$, when the system is initially prepared in $\ket{\psi (0)} = \ket{\uparrow \downarrow} \otimes \ket{\alpha}$, for $\Delta=0$ ($\gamma=\omega$: resonance), $\lambda_-/\omega=0.0001$ ($\lambda_1/\omega=0.8, ~ \lambda_2/\omega= 0.7999$), $\alpha=7$.}
\end{center}
\end{figure}

In this scenario we can analytically calculate the concurrence between the qubits which turns out to be:
\begin{equation}
    \mathcal{C}(t) = {1 \over 2} \left| 1 - \sum_n c_n^2 \cos(2 \sqrt{n} \tau_-) \right|,
\end{equation}
and is plotted in Fig. \ref{fig: Conc_JCM}.
The entanglement initially exhibits rapid oscillations due to the contribution of many frequency components. As time progresses, dephasing among the different $n$ terms reduces the concurrence, which settles to a constant value of $1/2$.
The resulting dynamics therefore feature the generation of entanglement that is not completely suppressed by the mode in the long-time limit, but instead persists in a partially reduced form.
{\color{black}
We see that the temporal evolution of entanglement also reveals a counterintuitive dynamical behavior.
In principle, strong qubit–mode coupling is expected to completely suppress the entanglement between the qubits due to the counter-rotating terms \cite{WangNJP13}.
Nevertheless, owing to the emergence of an effective weak-coupling regime governing the collective dynamics, a finite fraction of the entanglement can survive asymptotically.
We emphasize that this effect differs from the one reported in \cite{ManiscalcoPrl08}, which concerns the death and rebirth of entanglement in a two-qubit system coupled to a lossy reservoir. In that scenario, the qubit–bath interaction Hamiltonian contains only rotating terms, regardless of the coupling regime. In contrast, in our case the Jaynes–Cummings form of the effective Hamiltonian is derived analytically and rigorously justified on the basis of symmetry considerations.
}

When the two qubits couple equally to the cavity mode (i.e., $\lambda_1=\lambda_2$), the expectation values $\average{\sigma_1^z}$ and $\average{\sigma_1^x\sigma_2^x}$ remain constant (equal to 1 and 0, respectively).
This occurs because, under these conditions, the fictitious spin $b$, which represents the two qubits within the subspace $\mathcal{H}_b$, becomes decoupled from the field mode.
As a result, even though both qubits interact with the quantized field at equal coupling strengths, they are effectively decoupled from the mode within the subspace $\mathcal{H}_b$.
This means that the two qubits evolve collectively as if the cavity were absent.
Notably, this phenomenon is related to the fact that $\mathcal{H}_b$ becomes a decoherence-free subspace when $\lambda_1=\lambda_2$, also in presence of a collection of harmonic oscillators rather than a single mode \cite{GMNSV, GVSM}.

Thus, by initializing the two-qubit-mode system in the state given by Eq. \eqref{In State}, the characteristic time evolution of a JCM dynamics can be probed by measuring the single-qubit z-components $\average{\sigma_j^z}$ ($j=1,2$) which manifest the characteristic collapses and revivals of a JCM dynamics.
This is remarkably curious in our scenario, since the two qubits are strongly coupled to the cavity mode.
In the case corresponding to Fig. \ref{fig: mvs subdynamics b}, both qubits are indeed strongly coupled to the mode frequency, namely $\lambda_1/\gamma = 0.8, ~ \lambda_2/\gamma = 0.7999$ ($\omega=\gamma$).
Nevertheless, this is well explained by the fact that the dynamics of the two interacting qubits is ruled by an effective coupling parameter (see Eq. \eqref{H- new}) which results to be much lower than the other Hamiltonian parameters.

\begin{figure}[htp]
\begin{center}
\subfloat[][]{\includegraphics[width=0.22\textwidth]{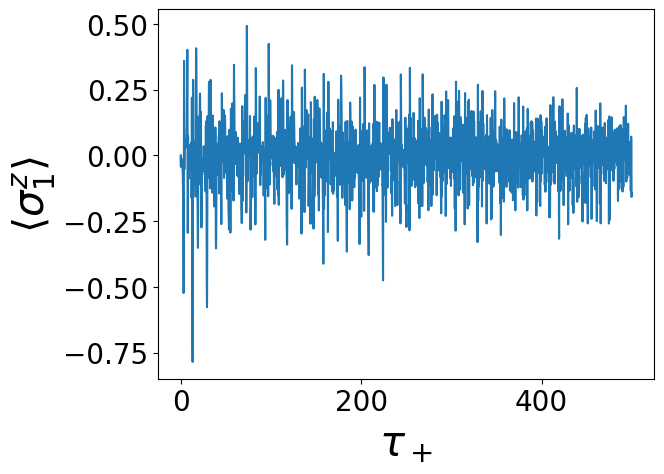}\label{fig: mv_z_num_strong}}
\qquad
\subfloat[][]{\includegraphics[width=0.22\textwidth]{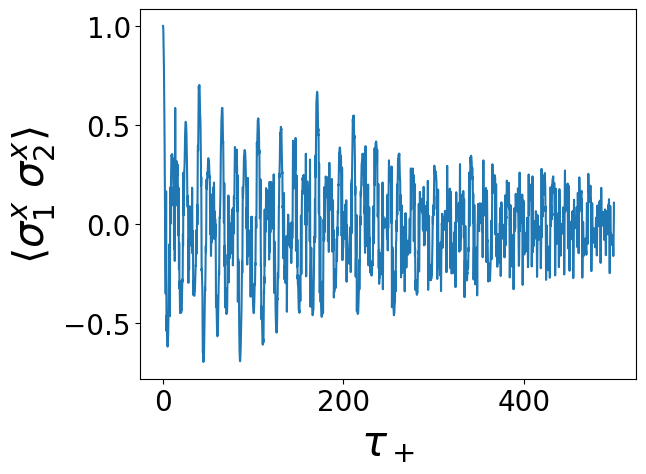}\label{fig: mv_xx_num_strong}}
\captionsetup{justification=raggedright,format=plain,skip=4pt}%
\caption{Time behavior, versus $\tau_+=\lambda_+ t$, of: (a) $\average{\sigma_1^z}$ (numerical), and (b) $\average{\sigma_1^x\sigma_2^x}$ (numerical), when the system is initially prepared in $\ket{\widetilde{\psi} (0)} = \ket{\uparrow \uparrow} \otimes \ket{\alpha}$, for $\Delta=0$ ($\gamma=\omega$: resonance), $\lambda_+/\omega=1.5999$ ($\lambda_1/\omega=0.8, ~ \lambda_2/\omega= 0.7999$), $\epsilon_1=\epsilon_2=\omega/2$, $\alpha=7$.}
\label{fig: mvs subdynamics a}
\end{center}
\end{figure}

In these conditions the other subdynamics results to be governed by the effective Hamiltonian $H_a$ which presents the effective coupling parameter $\lambda_+$.
The significantly different time behavior of both $\average{\sigma_1^z}$ and $\average{\sigma_1^x\sigma_2^x}$ from the ones seen before, when the system is initially prepared in the state
\begin{equation}
    \ket{\widetilde{\psi}(0)} = \ket{\uparrow \uparrow} \otimes \ket{\alpha},
\end{equation}
is shown in Fig. \ref{fig: mvs subdynamics a}.
In this case, unlike what happens in the subspace $b$, the emerging dynamics is characterized by an ultrastrong coupling ($\lambda_+=1.5999$), although the two qubits are just strongly coupled to the cavity mode.
In this instance it is thus present an opposite effect to the one highlighted before: the enhancement of the coupling between the two-qubit system and the quantized field.

These results highlight that in many-body physics, the dynamics of a quantum system is not strictly dictated by the interaction regime between the single qubits and the cavity mode.
This particular aspect is due to the fact that the dynamics typically associated with a particular interaction strength can be produced by effective parameters (derived from combinations of the actual Hamiltonian parameters).
This can leads to counter-intuitive outcomes, challenging the expectations based solely on the `real' physical parameters.
Consequently, in multi-qubit systems with varying parameters, new physical criteria can emerge to define the strong and weak coupling regimes, analogous to the adjustments seen in defining the thermodynamic limit \cite{GFMPSSV}.
It is important to note that this situation does not occur in systems described by the Dicke \cite{Dicke54} or Tavis-Cummings \cite{Tavis-Cummings68} models.
This is because, in those cases, the qubits within the system are not directly coupled, allowing the system to be interpreted as a collection of independent single-qubit QRMs and JCMs, respectively.
In contrast, in our case, the key distinguishing factor, from a dynamical perspective, is the qubit-qubit interaction.
The latter causes the two qubits to collectively behave as a single two-level system within the two invariant subspaces under actual Hamiltonians defined by effective parameters.

\textit{Two-Qutrit QRM.}
The circumstance of a JCM dynamics exhibited by two subsystems, although the latter are strongly coupled to a cavity mode, is not restricted to just the case of two interacting qubits.
In order to prove the wider generality of the peculiar effect seen above, let us consider the same model in Eq. \eqref{Hamiltonian} for two interacting qutrits:
\begin{equation} \label{Qutrits QRM}
\begin{aligned}
H =&
\Omega \Sigma_{1}^{z}+\Omega \Sigma_{2}^{z}
+\gamma_{x}\Sigma_{1}^{x}\Sigma_{2}^{x} + \omega a^\dagger a + 
\sum_{k=1}^2 \lambda_{k} \left( a^\dagger + a \right) \Sigma_k^z,
\end{aligned}
\end{equation}
where $\Sigma^l$ ($l=x,y,z$) are the Pauli matrices for a qutrit (spin-1).
%This model exhibits both level crossings and a second-order supperradiant QPT in the strong (qutrit-mode) coupling regime and when the ratio of characteristic frequency of the cavity mode to the energy splitting of the qutrits tends to zero: $\omega/\Omega \rightarrow 0$ \cite{GdCFPMV}.
This two-qutrit-mode model, similarly to the two-qubit case, possesses a conserved quantity, namely $\Sigma_{tot}^z=\Sigma_1^z + \Sigma_2^z$, leading to a family of dynamically invariant Hilbert subspaces \cite{GMIV, GNMV}.
For our purpose, we concentrate on the subspace corresponding to the vanishing eigenvalue of $\Sigma_{tot}^z$, spanned by the two-qutrit-mode states $\{ \ket{1,-1}, \ket{00}, \ket{-1,1} \} \otimes \ket{n}$, with $\{ \ket{1}, \ket{0}, \ket{-1} \} $ being the eigenstates of the single-qutrit operator $\Sigma^z$.
In such a subspace, analogously to the case of two qubits, an effective Hamiltonian of a single qutrit interacting with the boson field mode can be written:
\begin{equation}\label{H3}
H'_0=
\gamma \Sigma^{x} +
\omega a^\dagger a + 
2\lambda_- \left( a^\dagger + a \right) \Sigma^z.
\end{equation}
In this subspace the two qutrits behave thus as an effective three-level system.

\begin{figure}[htp]
\begin{center}
\subfloat[][]{\includegraphics[width=0.22\textwidth]{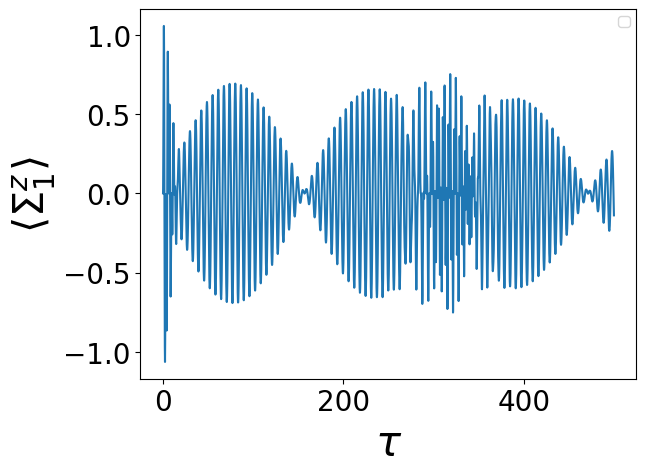}\label{fig: mv_z_qutrit}}
\qquad
\subfloat[][]{\includegraphics[width=0.22\textwidth]{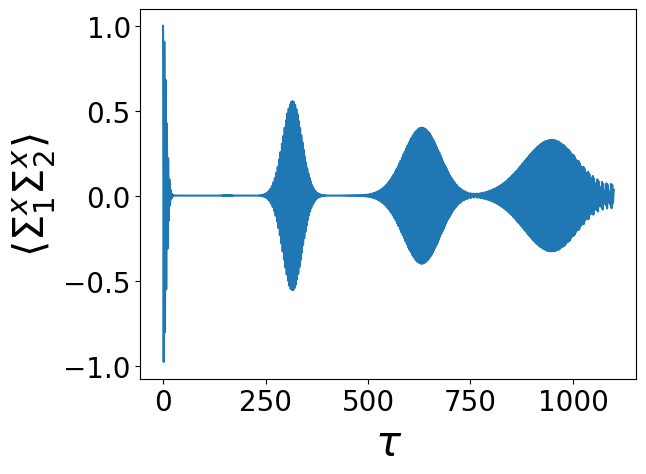}\label{fig: mv_xx_qutrit}}
\captionsetup{justification=raggedright,format=plain,skip=4pt}%
\caption{Time behavior, versus $\tau_-=\lambda_-t$, of: (a) $\average{\Sigma_1^z}$, and (b) $\average{\Sigma_1^x\Sigma_2^x}$, when the qutrit-mode system is initially prepared in $\ket{\Psi (0)} = (\ket{1-1}+\sqrt{2}\ket{00}+\ket{-11})/2 \otimes \ket{\alpha}$, for $\Delta=0$ ($\gamma=\omega$: resonance), $\lambda_-/\omega=0.05$, $\alpha=7$, $\tau = \omega t$.}
\label{fig: C&R qutrit}
\end{center}
\end{figure}

As observed in the case of the two qubits, we see that the effective spin-coupling $\lambda_-$, appearing in $H_0'$, depends on the difference between the coupling strengths of the two qutrits with the cavity mode.
Similarly in this scenario, if the two coupling parameters are nearly equal, i.e. $\lambda_1 \approx \lambda_2$, the effective interaction strength $\lambda_-$ can be considered much smaller than the qutrit-qutrit coupling $\gamma$ and the mode frequency $\omega$, despite the two qutrits are strongly coupled to the field mode.
Then, the counter-rotating terms can be neglected and the effective Hamiltonian $H_0'$ can be cast in the form of a JCM for a single three-level system, namely
\begin{equation}\label{H3 JCM}
\widetilde{H}'_0=
\gamma \Sigma^{z} +
\omega a^\dagger a - 
2\lambda_- \left( a^\dagger \Sigma^- + a \Sigma^+ \right).
\end{equation}
This implies that, by initializing the two qutrits in a linear combination of the three states defining this subspace, the system’s dynamics resemble that of a single three-level system weakly coupled (if $\lambda_1=\lambda_2$) to a single bosonic field mode.
Consequently, typical JCM dynamical phenomena, such as collapses and revivals, can occur even in the strong-coupling regime between each qutrit and the cavity.
This possibility is confirmed in Fig. \ref{fig: C&R qutrit} where the time behaviors of $\average{\Sigma_1^z}$ and $\average{\Sigma_1^x\Sigma_2^x}$ are shown.
Notably, when $\lambda_1=\lambda_2$, the two-qutrit system’s evolution (restricted to this subspace) becomes independent of the cavity mode.
In this case, indeed, the qutrits, collectively behaving as a single three-level system, effectively decouple from the field (since $\lambda_-=0$).

%\Ignore{
\textit{Qubit chain QRM.}
Let us consider now the following model
\begin{equation}
\begin{aligned}
H = \Omega \sum_k \hat{\sigma}_k^{z} + \gamma \bigotimes_{k} \hat{\sigma}_k^x +
\omega \hat{a}^\dagger \hat{a} + \sum_k \delta_k (\hat{a} + \hat{a}^\dagger) \hat{\sigma}_k^z
\end{aligned}
\end{equation}
where all the spin-qubits interact between each other through an $N$-wise interaction term and are all coupled to a quantized field mode.
Consider, without loss of generality, an even number $N$ of spins.

Basing on the exact treatment in Ref. \cite{GLSM}, the existence of $2^{N-1}$ dynamically invariant Hilbert subspaces can be established.
Notably, the interaction term involving the bosonic mode does not affect the symmetries of the Hamiltonian.
%Indeed, the spin-boson terms can be formally considered as fictitious $z$-field applied on the spin-qubits, more precisely, as a unique homogeneous field applied on the entire chain.
This circumstance stems from the existence of the $2^{N-1}$ constants of motion $\hat{\sigma}_i^z\hat{\sigma}_j^z$ (with $i \neq j$) \cite{GLSM}.
Within each of these invariant subspaces, a basis can be constructed as: $\{ \ket{e_k}, (\bigotimes_{j}\hat{\sigma}_j^x)\ket{e_k} \} \otimes \ket{n}$, where $\ket{e_k}$ represents a state of the standard $N$-qubit basis in the Hilbert space $\mathcal{H}$, and $(\bigotimes_{k}\hat{\sigma}_k^x)\ket{e_k}$ is the corresponding flipped state.
%For example, the states $\ket{\uparrow}^{\otimes(N-m)}\ket{\downarrow}^{\otimes m}$ and $\ket{\downarrow}^{\otimes(N-m)}\ket{\uparrow}^{\otimes m}$ illustrate this structure.

Each subdynamics is then governed by an effective Hamiltonian describing a two-level system interacting with a single cavity mode.
It means that the $N$-spin system behaves as an effective two-level system within each of the $2^{N-1}$ subdynamics.
The formal expressions of these distinct Hamiltonians can be derived analytically, with their structure depending on the nature of the states spanning the corresponding subspace.
More specifically, the form of each Hamiltonian is determined by the number of spin-up and spin-down components in the standard basis states.
By indicating with $m$ the number of spin-down, the Hamiltonian governing the dynamics in the subspaces spanned by states with $m$ spin-down, such as $\ket{\uparrow}^{\otimes(N-m)}\ket{\downarrow}^{\otimes m}$, reads
\begin{equation}
\begin{aligned}
    H_m = \Omega_m \hat{\sigma}^z + \gamma \hat{\sigma}^x +
    \sum_j \omega_j \hat{a}_j^\dagger \hat{a}_j + \delta_m (\hat{a} + \hat{a}^\dagger) \hat{\sigma}^z,
\end{aligned}
\end{equation}
with
\begin{equation}
\begin{aligned}
    \Omega_m = (N-2m) \omega,
    \qquad
    \delta_m = \delta_1+\sum_{i=2}^{N}\prod_{j=2}^{i} \sigma_{j}^z \delta_i.
\end{aligned}
\end{equation}
Both expressions depend on the number $m$ of spin-down.
Specifically, in the expression of $\delta_m$, $m$ of the $\sigma_j^z$ operators take the value $-1$, and $N-m$ take the value $+1$.
The simplified expression of $\Omega_m$, where the parameter $m$ appears explicitly, reflects the assumption that all qubits have identical splitting energies.
%The index $m$ ranges from 0 to $N/2$ since the flipped states, e.g. $(\bigotimes_{k}\hat{\sigma}_k^x)\ket{\uparrow}^{\otimes(N-m)}\ket{\downarrow}^{\otimes m}=\ket{\uparrow}^{\otimes m}\ket{\downarrow}^{\otimes (N-m)}$, must not be included in the count.
%These flipped states serve indeed as complementary partners to $\ket{\uparrow}^{\otimes(N-m)}\ket{\downarrow}^{\otimes m}$, together forming the basis of the corresponding subspace.

For our purposes, we focus on the case $m=N/2$.
It is crucial to note that there exist exactly $\binom{N-1}{m}$ subspaces corresponding to this configuration.
%Examples include the subspaces spanned by the pairs $\{ \ket{\uparrow}^{\otimes N/2} \ket{\downarrow}^{\otimes N/2}, \ket{\downarrow}^{\otimes N/2} \ket{\uparrow}^{\otimes N/2} \}$ or $\{ \ket{\uparrow \downarrow \uparrow \downarrow \dots \uparrow \downarrow}, \ket{\downarrow \uparrow \downarrow \uparrow \dots \downarrow \uparrow} \}$.
When $m=N/2$, if the coupling parameters $\delta_k$ of each qubit with the mode are nearly identical, the effective coupling $\delta_m$ becomes almost negligible.
Consequently, it can be assumed to be much smaller than $\gamma$ and $\omega$.
In this case, thus, the Hamiltonians $H_m$, governing the dynamics in these peculiar subspaces, can be approximated by a JCM Hamiltonian.
Therefore, by preparing the $N$ qubits in one of these states, the characteristic dynamical effects of the JCM can be observed by measuring $\average{\sigma_1^z}$ whose time behavior is identical to that shown in Fig. \ref{fig: mv_z_an}.
Further, it is worth noting that all these subspaces, spanned by pairs of standard basis states with an equal number of spins in the up and down states, exhibit effective decoupling from the mode (or the bath) when $\delta_k=\delta, ~ \forall m$.
Under this special condition, these subspaces become decoherence-free subspaces, meaning the $N$-qubit system (holistically behaving as a two-level system) evolves as if the cavity mode were absent.

Similarly to the case of two qubits, in the subspace spanned by the states $\{ \ket{\uparrow}^{\otimes(N)}, \ket{\downarrow}^{\otimes(N)}\}$, the opposite effect takes place.
Here, the effective coupling to the mode results indeed from the sum of all individual qubit couplings to the cavity mode.
The effective Hamiltonian ruling the dynamics in the correesponding subspace reads indeed
\begin{equation}
    H_{\{ \ket{\uparrow}^{\otimes(N)}, \ket{\downarrow}^{\otimes(N)}\}} = N \Omega \sigma^z + \gamma \sigma^x + N \delta (a + a^\dagger) \sigma^z,
\end{equation}
where, in this case, we have supposed the same interaction strength of each qubit to the mode: $\delta_k = \delta, ~ \forall k$.
Consequently, through this type of interaction among qubits, it is possible to achieve strong, ultra-strong, or even deep strong effective coupling, even when each qubit is only weakly coupled to the mode, simply by adjusting the number of qubits in the chain.
This is particularly noteworthy, as such interactions can be realized using both trapped ion systems \cite{Muller11} and superconducting circuits \cite{Mezzacapo14}.
%}

{\color{black}
Therefore, both the two-qutrit and the $N$-qubit chain QRMs exhibit the same counterintuitive behavior: the dynamical coupling regime displayed by the systems does not correspond directly to the qubit–mode interaction strength.
Instead, it depends crucially on the symmetries of the effective Hamiltonian governing the subdynamics associated with the chosen initial state.
This observation highlights how the distinction between strong and weak coupling in systems with many interacting qubits and modes can differ substantially from the standard picture and may become significantly more complex, as it depends on the specific physical characteristics of the system and the resulting mathematical structure of the Hamiltonian.
}

\textit{Conclusions.}
This work analyzes strong and weak coupling in many-spin systems interacting with a quantized field, showing that standard single-qubit QRM criteria do not directly apply.
The coupling regime must be redefined due to two key factors: (1) different physical conditions in many-body systems and (2) collective dynamics that create an effective coupling distinct from individual spin-field interactions.

%Invariant subspaces, arising from system symmetries, lead to effective Hamiltonians with redefined parameters.
%Consequently, the coupling regime depends on spin-spin coupling relative to mode frequency rather than the conventional spin splitting-to-mode frequency ratio.
%This is demonstrated in two-qubit, two-qutrit, and $N$-qubit chain systems.

Importantly, the effective coupling can be significantly weaker than the original parameters, altering system dynamics.
Collapse and revival, typical of weak coupling, can occur even when individual spins strongly interact with the field.
Conversely, the opposite effect can occur: collective dynamics can be characterized by a stronger coupling than the actual one between spins and field.

These results highlight the complexity of defining coupling regimes in many-body systems, where effective Hamiltonians dictate dynamics, leading to unexpected and counterintuitive effects.

\section*{Acknowledgements}
This work was supported by the PNRR MUR project PE0000023-NQSTI.
GF and EP acknowledge the QuantERA grant SiUCs (Grant No. 731473) and the grant Pia.Ce.Ri.-UNICT, project Q-ICT.

\bibliography{biblio_qrm.bib}
%\printbibliography

\end{document}